\documentstyle[aps,multicol]{revtex}

\begin{document}
\draft
\title{Fault-tolerant quantum computation with long-range correlated noise}
\author{Dorit Aharonov,$^1$ Alexei Kitaev,$^{2,3}$ and John Preskill$^2$}
\address{
$^1$ School of Computer Science and Engineering, Hebrew University, Jerusalem, Israel\\
$^2$ Institute for Quantum Information, California Institute of Technology, 
Pasadena, CA 91125, USA\\
$^3$ Microsoft Research, One Microsoft Way, Redmond, WA 98052, USA 
}
\maketitle
\begin{abstract}
We prove a new version of the quantum accuracy threshold theorem that applies to non-Markovian noise with algebraically decaying spatial correlations. We consider noise in a quantum computer arising from a perturbation that acts collectively on pairs of qubits and on the environment, and we show that an arbitrarily long quantum computation can be executed with high reliability in $D$ spatial dimensions, if the perturbation is sufficiently weak and decays with the distance $r$ between the qubits faster than $1/r^D$. 


\end{abstract}
\pacs{PACS numbers: 03.67.Pp}
 
\begin{multicols}{2}

The theory of quantum fault tolerance \cite{shor_ft} establishes that quantum information, if cleverly encoded, can be protected from damage and processed reliably with imperfect equipment. One noteworthy achievement of this theory is the
{\em quantum threshold theorem}  \cite{ben-or,kitaev_threshold,knill-laflamme,aliferis,reichardt}, which asserts that an arbitrarily long quantum computation can be executed with high reliability, provided that the noise afflicting the computer's hardware is weaker than a certain critical value, the {\em accuracy threshold}. 

Early proofs of the threshold theorem \cite{ben-or,kitaev_threshold,knill-laflamme} assumed that the noise is {\em Markovian}, that each quantum gate in the noisy circuit is a trace-preserving completely positive map that approximates the ideal gate. 
Each gate in this noise model can be realized as a unitary transformation that acts jointly on a set of the qubits in the computer (which we will call {\em system qubits}) and on the environment (the {\em bath variables}), but where it is assumed that the bath has no memory --- the state of the bath is refreshed after every gate. In \cite{terhal} the theorem was extended to a non-Markovian noise model of a restricted type. It was assumed that each qubit $Q_i$ has a {\em local} bath $E_i$, and that $E_i$ and $E_j$ interact only when a quantum gate acts on the corresponding pair of qubits. In \cite{aliferis}, the noise model was relaxed further by allowing nonlocal interactions among the bath variables, but it was still assumed that a pair of system qubits interact directly only when a quantum gate acts on that pair. In this paper, we will show that the threshold theorem applies even if there are long-range interactions among the system qubits that are ``always on.'' 

The noise model we will consider can be formulated in terms of a time-dependent Hamiltonian $H$ that governs the joint evolution of the system and the bath. We may express $H$ as
\begin{equation}
H=H_S+H_B +H_{SB}~,
\end{equation}
where $H_S$ is the time-dependent Hamiltonian of the system that realizes the ideal quantum circuit, $H_B$ is an arbitrary Hamiltonian of the bath, and  $H_{SB}$ couples the system to the bath. In \cite{aliferis} it was assumed that the system-bath coupling could be expressed as 
\begin{equation}
H_{SB}=\sum_a H_{SB,a}~;
\end{equation} 
here $a$ labels a circuit ``location'' during a particular time step of the computation --- either a single qubit at a location where a one-qubit gate acts in the ideal circuit (which could be the identity gate in the case of a ``resting'' qubit), or a pair of qubits at a location where a two-qubit gate acts on the pair in the ideal circuit. This coupling of the system qubits to one another is ``short range'' in the sense that two system qubits interact directly with one another only if the ideal Hamiltonian $H_S$ also couples those two data qubits during the same time step --- hence we will refer to it as the {\em short-range noise model}. For short-range noise it was shown in \cite{aliferis} that there is a positive constant $\varepsilon_0$ such that an arbitrarily long ideal quantum computation can be accurately simulated using the noisy gates (with reasonable overhead) provided that at each time and at each location
\begin{equation}
\label{short-bound}
\parallel H_{SB,a}\parallel t_0 < \varepsilon_0~,
\end{equation}
where $t_0$ is the time needed to execute a quantum gate and $\parallel \cdot \parallel$ denotes the sup operator norm. 

In this paper we will consider a system-bath coupling of the form
\begin{equation}
H_{SB}=\sum_{<ij>} H_{ij}~.
\end{equation}
where the sum is over all pairs of system qubits; $H_{ij}$ acts collectively on the pair of qubits $<ij>$ and on the bath. (By convention, $H_{ij}=H_{ji}= H_{<ij>}$.) Since any pair of system qubits can interact, we call this the {\em long-range noise model.} We will show that for long-range noise there is a positive constant $\eta_0$ such that an arbitrarily long ideal quantum computation can be accurately and efficiently simulated provided that for each qubit $i$ and at each time
\begin{equation} 
\label{H-sum}
\sum_j \parallel H_{ij}\parallel t_0 < \eta_0~.
\end{equation}
We note that if the interaction between system qubits decays algebraically so that
\begin{equation}
\label{interaction-decay}
\parallel H_{ij} \parallel < \delta/|i-j|^z~,
\end{equation}
where $|i-j|$ denotes the distance between qubits $i$ and $j$, then the sum in eq.~(\ref{H-sum}) converges in the limit of an infinite $D$-dimensional system provided that $z>D$. Thus, our result shows that fault-tolerant quantum computation is possible for $\delta$ sufficiently small.

As in \cite{terhal,aliferis}, we will analyze the noisy circuit using a ``fine-grained fault-path expansion.'' We divide the working period $t_0$ of a gate into $t_0/\Delta \gg 1$ intervals each of width $\Delta$; for short-range noise the time evolution operator for the time interval $(t,t+\Delta)$ can then be expressed as
\begin{eqnarray}
U(t+\Delta,t)&&\approx e^{-i\Delta H}\nonumber\\
&&\approx e^{-i\Delta H_S}e^{-i\Delta H_B}\prod_a\left(I_{SB}- i\Delta H_{SB,a}\right)
\end{eqnarray}
since for $\Delta$ sufficiently small terms higher order in $\Delta$ can be safely neglected. The action $U_{\rm circuit}$ on the system and bath of a noisy quantum circuit with altogether $N$ time steps can be expressed as a product of $Nt_0/\Delta$ such evolution operators. The fine-grained fault-path expansion of the noisy circuit is constructed by expressing $U_{\rm circuit}$ as a sum of monomials, where in each term either $I_{SB}$ or $-i\Delta H_{SB,a}$ is inserted at each ``micro-location'' with duration $\Delta$. A micro-location where $-i\Delta H_{SB,a}$ is inserted is called a ``micro-fault.''
A noisy gate or ``macro-location'' in the circuit consists of $t_0/\Delta$ consecutive micro-locations, and the macro-location simulates the ideal gate perfectly if it contains no micro-faults. Therefore, we say that a macro-location is faulty (is a ``macro-fault'') if it contains one or more micro-faults. By grouping together terms in the fine-grained fault-path expansion we obtain a coarse-grained fault-path expansion such that each term in the coarse-grained expansion is the sum of all terms in the fine-grained expansion with macro-faults occuring at a specified set of macro-locations. 

The strength of the noise in the circuit can be quantified by the norm of the sum $E({\cal I}_1)$ over all the terms in the fine-grained expansion such that a particular specified macro-location ${\cal I}_1$ is faulty (where no restriction is placed on the noise at other macro-locations). We can organize this sum by noting that there must be an earliest micro-fault in the macro-location which can occur at any of the $t_0/\Delta$ micro-locations; summing over all fine-grained fault paths with a fixed earliest micro-fault is equivalent to inserting $I_{SB}$ at the micro-locations prior to the earliest micro-fault and inserting $e^{-i\Delta H_{SB,a}}$ at the later micro-locations. Since $\parallel e^{-i\Delta H_{SB,a}}\parallel=1$ and the norm of a product is bounded above by a product of norms, the norm of the sum over all fine-grained paths with a fixed earliest micro-fault is no larger than $\Delta \parallel H_{SB,a}\parallel$. Now summing over the $t_0/\Delta$ possible micro-locations for the earliest micro-fault, and noting that the norm of a sum is bounded above by a sum of norms, we have
\begin{equation}
\parallel E({\cal I}_1) \parallel~ \le ~ \varepsilon \equiv {\rm Max} \parallel H_{SB,a}\parallel t_0 ~,
\end{equation}
where the maximum is with respect to all $a$ and all times. By similar reasoning, if $E({\cal I}_r)$ denotes the sum over all fine-grained fault paths such that faults occur at a particular set ${\cal I}_r$ of $r$ macro-locations (where no restriction is placed on the noise at other macro-locations), then
\begin{equation}
\parallel E({\cal I}_r)) \parallel ~\le~ \varepsilon^r~.
\label{local-noise-defn}
\end{equation}
If eq.~(\ref{local-noise-defn}) is satisfied, we say that the noise is ``local with strength $\varepsilon$.'' Note that $\varepsilon$ is determined only by the system-bath coupling $H_{SB}$; no restriction need be imposed on the bath Hamiltonian $H_B$.

Thus we see that the short-range noise model obeys the local noise condition eq.~(\ref{local-noise-defn}). For a quantum computer subjected to local noise, the effectiveness of a recursive fault-tolerant simulation of an ideal circuit was analyzed in \cite{aliferis}. In a level-1 simulation, each gate of an ideal circuit is replaced by a 1-gadget, a circuit that simulates the gate acting on quantum information protected by a quantum error-correcting code. In a level-$k$ simulation, each gate of an ideal circuit is replaced by a $k$-gadget, which  is constructed by replacing each gate in a $(k{-}1)$-gadget by a 1-gadget. If the code can correct $t$ errors, then a 1-gadget is said to be {\em bad} if it contains $t+1$ or more faulty macro-locations, and a $k$-gadget is said to be bad if it contains $t+1$ or more bad $(k{-}1)$-locations. 

If the 1-gadgets are properly constructed, then it can be shown that for a level-1 simulation, the norm of the sum $E({\cal I}_r^{(1)})$ over fault-paths such that the 1-gadgets are bad at a specified set ${\cal I}_r^{(1)}$ of $r$ level-1 locations satisfies
\begin{equation}
\parallel E({\cal I}_r^{(1)}) \parallel ~\le~ \left(\varepsilon^{(1)}\right)^r~,
\end{equation}
where 
\begin{equation}
\varepsilon^{(1)}=C {A\choose t+1}\varepsilon^{t+1}~;
\end{equation}
here $A$ is the maximal number of macro-locations contained in any 1-gadget, and $C$ is a constant of order 1. 
Furthermore, by an inductive argument we find that for a level-$k$ simulation, the norm of the sum $E({\cal I}_r^{(k)})$ over fault-paths such that the $k$-gadgets are bad at a specified set ${\cal I}_r^{(k)}$ of $r$ level-$k$ locations satisfies
\begin{equation}
\parallel E({\cal I}_r^{(k)}) \parallel ~\le~  \left(\varepsilon^{(k)}\right)^r~,
\end{equation}
where 
\begin{eqnarray}
\varepsilon^{(k)}= C {A\choose t+1}\left(\varepsilon^{(k{-}1)}\right)^{t+1}= \varepsilon_0 \left(\varepsilon/\varepsilon_0\right)^{(t+1)^k}~,
\end{eqnarray}
and
\begin{eqnarray}
\varepsilon_0=\left(C{A\choose t+1}\right)^{-1/t}~.
\end{eqnarray}
Thus, for $\varepsilon < \varepsilon_0$, the norm drops steeply as the level $k$ increases, and it can be shown that an ideal circuit with $L$ gates can be simulated with fixed accuracy and an overhead cost that is polylogarithmic in $L$. This is the quantum threshold theorem for local noise \cite{aliferis}.

To establish a threshold theorem for the long-range noise model, then, it will suffice to show that the local noise condition eq.~(\ref{local-noise-defn}) holds for the long-range noise, just as it does for short-range noise. For now, suppose that all of the gates performed at the $r$ specified faulty macro-locations are single-qubit gates. If these macro-locations all occur at {\em different} times, then our earlier analysis of short-range noise needs little modification. Now we express the time evolution operator for time interval $(t,t+\Delta)$ as
\begin{eqnarray}
U(t+\Delta,t) \approx e^{-i\Delta H_S}e^{-i\Delta H_B}\prod_{<ij>}\left(I- i\Delta H_{ij}\right)
\end{eqnarray}
and again expand $U_{\rm circuit}$ to generate a fine-grained fault-path expansion. Considering a fixed faulty macro-location ${\cal I}_1$ where the ideal gate acts on qubit $i$, again there must be an earliest micro-fault that acts on qubit $i$ in that macro-location, which can be at any of $t_0/\Delta$ intervals; we insert 
$-i\Delta \left(\sum_j H_{ij}\right)$ at that earliest micro-fault, $I$ at each earlier micro-location, and $e^{-i\Delta H_{SB}}$ at all later time intervals. Then summing over the position of the earliest micro-fault we find
\begin{equation}
\parallel E({\cal I}_1)\parallel ~\le~ \eta \equiv {\rm Max} \left(\sum_j \parallel H_{ij}\parallel\right) t_0~,
\end{equation}
where the maximum is with respect to all $i$ and all times. Similarly, if macro-faults occur at a set of ${\cal I}_r$ of  $r$ macro-locations, where all the macro-locations are at different time steps, then by summing independently over the earliest micro-fault at each macro-location, we find
\begin{equation}
\parallel E({\cal I}_r)\parallel ~\le~ \eta^r~.
\end{equation}

But what if many macro-faults occur during the same working period? The long-range noise model allows faults occuring in the same time step to be correlated. The worst case is when the $r$ specified faulty macro-locations in the set ${\cal I}_r$ are all in the same time step. In that case, if $i$ and $j$ are two of the qubits acted upon by the $r$ faults, then $H_{ij}$ can account for the two of the faults. 

We again organize the sum over fine-grained fault paths by considering the earliest micro-fault that strikes each of the $r$ qubits. For a particular fine-grained fault path, if $i,j\in {\cal I}_r$ and  $H_{ij}$ is the first perturbation to act on both of qubits $i$ and $j$, then we say that the pair $ij$ is ``contracted.'' If $H_{ij}$ is the first perturbation to act on qubit $i$, where either $j\not\in {\cal I}_r$ or $j\in {\cal I}_r$ has already been struck earlier by another perturbation, then we say that qubit $i$ is ``uncontracted.'' By definition, then, the contracted pairs are disjoint. 

We group the fine-grained fault paths into families, where in each family the time of the earliest micro-fault acting on each qubit in ${\cal I}_r$ is specified, as well as the pair of qubits acted upon by that fault. If is clear that the norm of the sum of fine-grained fault paths in a family can be bounded by a product --- with a factor of $\Delta\parallel H_{ij}\parallel$ arising from a fault that acts on the pair $<ij>$; the only problem is to count the families. 

In each family, a specified set of $k$ pairs are contracted, where $k$ ranges from $0$ to $r/2$ for $r$ even, or from 0 to $(r-1)/2$ for $r$ odd. Each contraction can occur in any of the $t_0/\Delta$ time steps, which converts the factor of $\Delta$ to $t_0$. For an uncontracted qubit $i$, we sum over the qubit $j$ it could be paired with by the perturbation $H_{ij}$ --- the sum runs over both $j\in {\cal I}_r$ and $j\not\in{\cal I}_r$. For $j\not\in {\cal I}_r$, this micro-fault can occur at any time, but for $j\in {\cal I}_r$ that might not be true. If we slide $H_{ij}$ forward in time to before the first micro-fault acting on qubit $i$, then we are considering a family in which $i$ and $j$ {\em are} contracted. But that is all right; if we allow $H_{ij}$ to occur at any time, we are only overestimating the norm of the sum over fault paths; so here again we have an upper bound by converting $\Delta$ to $t_0$, and summing over $j$ gives a factor of $\eta$ for each uncontracted qubit. 

Let's suppose that $r=2n$ is even. The norm $S_k$ of the sum over fine-grained fault paths with $k$ contractions can be bounded above:
\begin{eqnarray}
S_k&&~\le  \sum_{\rm contractions}\eta^{2n-2k}\prod_{\ell=1}^k\left( \parallel H_{i_\ell,j_\ell}\parallel t_0\right)\nonumber\\
&&\le\frac{1}{2^k k!}\left( \sum_{i,j}\parallel H_{ij}\parallel t_0\right)^k \eta^{2n-2k}\nonumber\\
&&\le\frac{1}{2^k k!}\left( 2n\eta\right)^k \eta^{2n-2k}~.
\end{eqnarray}
(Note that the sum of $\parallel H_{ij}\parallel $ over all $i$ and $j$, raised to the power $k$, contains a term bounding the contribution from each contraction $2^k k!$ times, and also contains other positive terms.) Thus we have
\begin{equation}
S_k \le \frac{1}{(k/e)^k} n^k \eta^{2n-k}= e^k\left({n\over k}\right)^k\eta^{2n-k}~.
\end{equation}
To bound the sum over $k$ we use 
\begin{equation} 
\left({n\over k}\right)^k = \left(\left({k\over n}\right)^{-k/n}\right)^n \le \left(e^{1/ e}\right)^n
\end{equation}
and $n+1\le e^n$ to obtain
\begin{eqnarray}
\parallel E({\cal I}_r) \parallel =\sum_{k=0}^n S_k ~&&\le~ (n+1)e^n e^{n/e}\eta^n\nonumber\\
&& \le \left( {e^{2+1/e}} \eta\right)^{n} = (\eta')^{r}~,
\end{eqnarray}
where
\begin{equation}
\eta'=e^{1+1/2e}\sqrt{\eta}~.
\end{equation}
Thus, we have shown that, even when all of the $r=2n$ specified faults occur during the same working period, the norm of the sum over fault paths is bounded by $\eta'^{r}$, where $\eta'=O(\sqrt{\eta})$. This bound also applies when $r$ is odd.

The same bound applies to two-qubit gates, except that we need to replace $\eta$ by $2\eta$. A two qubit gate has a fault if the perturbation hits either one of the two qubits (and after one of the qubits is damaged, we don't care what happens to the other one). The first perturbation to damage the pair of qubits $i$ and $j$ might act on the pair $im$ ($m\ne j$), on the pair $mj$ ($m\ne i$), or on the pair $ij$. So the norm of the sum over fault paths that damage one particular two-qubit macro-location can be bounded above by
\begin{equation}
\left(\sum_{m\ne j}\parallel H_{im}\parallel + \sum_{m\ne i}\parallel H_{mj}\parallel+\parallel H_{ij}\parallel\right)t_0\le 2\eta~.
\end{equation}
When we consider summing over the fault paths with many faults in the same time step, where some of the faults occur at two-qubit macro-locations, now the ``uncontracted'' contributions include the case where a single perturbation damages both of the qubits at a single macro-location; if on the other hand a two-qubit macro-location is contracted with another macro-location, the perturbation can act on either qubit at the two-qubit macro-location. Both types of contributions are taken into account by replacing $\eta$ by $2\eta$. 

The conclusion is that ``long-range noise''  is really a type of ``local noise'' obeying eq.~(\ref{local-noise-defn}), where
\begin{equation}
\varepsilon = e^{1+1/2e}\sqrt{2 \eta}~.
\end{equation}
therefore the quantum accuracy threshold theorem proven in \cite{aliferis} is applicable. Note that we can use similar reasoning for many-body interactions. For example, if the system-bath interaction is a sum of three-body terms, then
\begin{equation}
\eta=  \left({\rm Max}\sum_{jk}\parallel H_{ijk}\parallel\right)t_0~,
\end{equation}
and $\varepsilon= O(\eta^{1/3})$. 

Actually, even the ``short-range'' noise model includes long-range spatial correlations among the faults.  If each system qubit couples individually to the bath with strength $\lambda$, but the self-interactions of the bath may be nonlocal, then a coupling between a pair of system qubits of order $\lambda^2$ can be induced in second-order perturbation theory, and $p$-body terms of order $\lambda^p$ are generated in higher order. Nevertheless, the ``long-range'' noise model is more general, because there are interactions among the system qubits that are allowed in the long-range model but cannot be simulated by a local coupling of the system to the bath. Furthermore, by working directly with a system-bath Hamiltonian that includes long-range interactions among the system qubits, we can see more easily how the rate of decay of the interactions with distance enters into the analysis, as in eq.~(\ref{interaction-decay}).

Using methods described in \cite{aliferis}, we can obtain a lower bound on the threshold noise strength $\varepsilon_0\approx 10^{-5}$, or $\eta_0\approx 10^{-10}$ for the case of two-body interactions. This estimate is discouraging, but it is probably much too pessimistic, for several reasons. In the analysis we allow the bad fault paths to add together with a common phase, so that they interfere constructively. Under a more plausible scenario in which the phases of distinct fault paths are only weakly correlated, the norm of the sum is significantly smaller, so that much stronger noise can be tolerated in practice. Furthermore, by following sound engineering principles, for example by spatially separating qubits belonging to the same code block, we can improve the effectiveness of fault-tolerant protocols. We also note that more clever gadget design could significantly improve estimates of the threshold as suggested for example in \cite{knill}, and in any case the actual value of the threshold is likely to be notably higher than can be established by rigorous arguments. The noteworthy point is that we have shown that sufficiently weak long-range interactions among the qubits (and interactions with a nonlocal  bath that can store information for an indefinitely long time) are compatible in principle with fault-tolerant quantum computation.

We thank Daniel Gottesman for helpful comments. This work has been supported in part by DoE under Grant No. DE-FG03-92-ER40701, NSF under Grant No. PHY-0456720, ARO under Grants No. W911NF-04-1-0236, W911NF-05-1-0294 and DAAD19-00-1-0374, ISF under Grants No. 032-9739 and 039-7549, the US Army under Grant No. 030-7657, and the Yigal Alon Scholarship under Grant No. 033-7233.

\end{multicols}
\end{document}